\documentclass[reprint,amsmath,amssymb,aps,nofootinbib,]{revtex4-2}

\usepackage{graphics}
\usepackage{amssymb}
\usepackage{amsmath}
\usepackage{amsfonts}
\usepackage{amssymb}
\usepackage{listings}
\usepackage{epstopdf}
\usepackage{listings}
\usepackage{epstopdf}
\usepackage{xcolor}
\usepackage{graphicx}
\usepackage[utf8]{inputenc}
\usepackage{here}
\usepackage{float flt}
\usepackage{enumitem}
\usepackage{booktabs}
\usepackage[toc]{appendix}
\usepackage{multirow}
\usepackage{ulem}
\newcommand {\E} {\varepsilon}
\newcommand {\Om} {\Omega}

\newcommand {\ee} {{\rm e}}

\renewcommand {\d} {{\rm d}}

\newcommand {\calA} {{\cal A}}
\newcommand {\calB} {\mathcal{B}}
\newcommand {\calC} {\mathcal{C}}
\newcommand {\calD} {{\cal D}}

\begin{document}

\title{Multiple scattering of 855 MeV electrons in
amorphous and crystalline silicon: simulations versus experiment}

\author{Germ\'an Rojas-Lorenzo}
\author{Jes\'us Rubayo-Soneira}
\author{Maykel M\'arquez-Mijares}
\affiliation{Instituto Superior de Tecnolog\'i{}as y Ciencias Aplicadas, Universidad de La Habana (InSTEC-UH),  Ave. Salvador Allende 1110, Plaza de la Revoluci\'on, Habana 10400, Cuba.}

\author{Andrei V. Korol}
\author{Andrey V. Solov'yov}
\affiliation{MBN Research Center, Altenhöferallee 3, 60438 Frankfurt am Main, Germany.}
\begin{abstract}
The angular distribution function of multiple scattering
experienced by 855 MeV electrons passing through an
amorphous silicon plate and an oriented silicon crystal
has been studied by means of relativistic molecular dynamics
simulations using two types of the potentials that describe
electron-atom interaction.
The differences in the angular distributions of the beam particles
in both media are analysed.
The results obtained are compared to the experimental data and to the
results of Monte Carlo simulations.
\end{abstract}
 
\maketitle

%%%%%%%%%%%%%%%%%%%%%%
\section{Introduction}
\label{intro}

In the 1960s, Lindhard published a comprehensive theoretical study
on the influence of crystal structure on the motion of energetic
charged particles \cite{Lindhard}.
He revealed and described directional effects for the projectiles
incident on oriented crystals which lead to the governed motion
due to inhomogeneity, anisotropy and lack of randomness in the
medium.
By introducing the concept of the continuous potential of an atomic
string or plane in a crystal Lindhard explained the channeling phenomenon
(i.e. the guided motion of charged projectiles in the vicinity of major
crystallographic axis or planes) that had been detected earlier in the
experiments on propagation lengths for heavy ions in aluminum
\cite{DaviesEtAl1960b}.
Since then studies of the channeling phenomenon, or more generally,
of passage of ultra-relativistic particles through oriented crystals
of different geometry (linear, bent, periodically bent crystals)
have expanded to form a broad field of theoretical and experimental
research (see, e.g., \cite{UggerhojRPM}) as well as of a number
of applications.
The latter include, for example, beam manipulation, --
steering, focusing, collimation, splitting, and extraction,
by means of crystals (see, e.g.,
Refs. \cite{BiryukovChesnokovKotovBook,SecondEdition,%
MazzolariEtAl:EPJC_v78_720_2018,%
DabagovGladkikh:RadPhysChem_v154_p3_2018,%
Scandale_EtAl-PR-AB_v21_014702_2018}
and references therein).
Another promising potential application concerns realisation of novel
intensive gamma-ray light sources operating at photon energies within
MeV-GeV range that can be constructed by exposing oriented crystals to beams of ultra-relativistic charged projectiles
\cite{CLS-book_2022,SushkoEtAl:EPJ_v76_166_2022,%
AVK-AVS:NIMB_v537_p1_2023}.

Regularity in positions of crystal constituents modifies multiple
scattering and radiative processes for a charged projectile particle
as compared to its passage through an amorphous medium
\cite{AkhiezerShulga:SovPhysUsp_v30_p197_1987}.
This occurs for the channeling particles as well as for those
experiencing the over-barrier motion, i.e. moving across crystal
planes / axes.

Recently, an experimental observation was reported
\cite{Scandale_EtAl:EPJC_v79_99_2019} of the reduction of multiple
scattering for high-energy positively charged particles
experiencing planar channeling in single crystals.
The measurements have shown that the root-mean-square planar angle
of multiple scattering in the plane normal to the plane of the
channeling is less than that for the over-barrier particles in the same
crystal.
The explanation presented refers to the specific feature of the
channeling phenomenon of positively charged projectiles.
Their interaction with the crystal atoms is repulsive, therefore,
in the planar regime they channel in between two adjacent planes.
If the amplitude of the channeling oscillations is smaller than
$d/2-a_{\rm TF}$ ($d$ is the interplanar distance and $a_{\rm TF}$
is the Thomas-Fermi radius of the atom) then channeling occurs in the
spatial region with the electron volume density less (on average)
than in the amorphous medium.
As a result, such particles can penetrate large distance into a crystal
and exit from the crystal at relatively small scattering angle.

In another recent experiment \cite{MazzolariEtAl:EPJC_v80_63_2020}
carried out at
the  Mainz Mikrotron (MAMI) facility the observation has been made of the
reduction of multiple scattering angle of 855 MeV electrons
incident at a small angle with respect to the $\langle 001 \rangle$
axis in a crystalline silicon as compared to that in an
amorphous target.
It is indicated in the cited paper that a projectile particle
while moving in a crystalline environment can interact
either with individual atoms as in an amorphous medium
(incoherent scattering) or with atoms arranged in
strings or in planes (coherent scattering).
Section 2 in the paper provides
quantitative analysis of the conditions to be met for
the incident angle of particle's momentum ${\vec p}_0$
with respect to the
axis (plane) that facilitate the coherent scattering and,
as a result, lead to the decrease in the scattering angle.
The effect of coherent scattering suppression (CSS) that
occur at certain crystal orientations has been observed experimentally
as well as modelled via Monte Carlo simulations
\cite{MazzolariEtAl:EPJC_v80_63_2020}.

In this paper we present an independent analysis of the multiple
scattering process of electrons in the crystalline and
amorphous silicon targets.
The analysis is based on the results of the relativistic classical
molecular dynamics simulations performed by means of the
\textsc{MBN Explorer} package
\cite{MBN_Explorer_2012,MBNExplorer_Book,MBN_Studio_2019,mbn-explorer-software}.
The parameters of the targets as well as the characteristic of the
electron beam (energy, angular divergence and orientation with respect to
the crystal) used in the simulations
correspond to those used in the experiment
\cite{MazzolariEtAl:EPJC_v80_63_2020} (see Section \ref{Theory} for
details).
In Section \ref{Results} the numerical results obtained are compared with
the experimental data collected at MAMI.
In Section \ref{Conclusion}, some conclusions of this work are
summarized and
future perspectives are presented.

%%%%%%%%%%%%%%%%%%%%%%%%%%%%%%%%%
\section{Methodology \label{Theory}}

A number of numerical tools have been developed within
frameworks of different theoretical approaches
to simulate various phenomena occurring during passage of
ultra-relativistic projectiles through oriented crystals and
amorphous solids
(see Ref. \cite{KorolSushkoSolovyov:EPJD_v75_p107_2021} for
and overview of the most recent ones).
In many cases, a rigorous description can be achieved
within the classical formalism by considering particles'
trajectories.
For numerical modeling of channeling and related phenomena
beyond the continuous potential framework we utilize the
multi-purpose computer package \textsc{MBN Explorer}
\cite{MBN_Explorer_2012,MBN_ChannelingPaper_2013,%
MBNExplorer_Book,mbn-explorer-software}
and a supplementary special multitask software toolkit
\textsc{MBNStudio} \cite{MBN_Studio_2019}.
Developed originally as a universal program for
investigating structure and dynamics of molecular systems
at various spatial scales, ranging from atomistic up to macroscopic,
the package allows for a computation of relativistic projectile
motion in various environments including
crystalline structures \cite{MBN_ChannelingPaper_2013}.
The simulation procedure accounts for the interaction of a
projectile particle with all atoms in the environment.
A number of various interatomic potentials implemented in
the package enables rigorous simulations of different media.
The algorithms implemented allow for modelling passage of
the particles over macroscopic distance with atomistic accuracy.
The package serves as a powerful numerical tool to explore
relativistic dynamics in crystals and amorphous systems.
Its efficiency and reliability have been evaluated for
channeling of ultra-relativistic projectiles within the
energy range of sub-GeV to tens of GeV in straight, bent,
periodically bent crystals, and amorphous medium
(see the review paper \cite{KorolSushkoSolovyov:EPJD_v75_p107_2021}
and references therein).

The simulation procedure of the motion an ultra-relati\-vis\-tic
particle of energy $\E$, charge $q$ and mass $m$ is carried out
within the framework classical relativistic mechanics.
\textsc{MBN Explorer} integrates relativistic equations of motion that are written as follows:
\begin{eqnarray}
\dot{\vec r} = {\vec v},
\qquad
\dot{\vec v} =
{1 \over m \gamma}
\left(
{\vec F} -
{{\vec v} \left({\vec v} \cdot {\vec F}\right)\over c^2}
\right).
\label{eq:01} %{Equations:eq.01}
\end{eqnarray}
Here ${\vec r}=\vec r(t)$ and ${\vec v}=\vec v(t)$ stand for
the particle's position vector and velocity at instant $t$,
$\gamma = \E/mc^2$
is the relativistic Lorentz factor.
The force $\vec F=-q \partial U({\vec r})/\partial {\vec r}$
 acting on the projectile is due to electrostatic field
created by atoms of the medium.

At each integration step the electrostatic potential
$U({\vec r})$ is computed as the sum of potentials $U_{\rm at}$
of individual atoms
\begin{equation}
 U({\vec r})=  \sum_{j} U_{\rm at}(| {\vec r} - {\vec R_j}|)\,.
\label{eq:02}
\end{equation}
Here ${\vec R}_j$ denotes a position vector of the $j$th atom.
The code allows for the evaluation of atomic potentials
using approximations proposed by Moli\`{e}re \cite{Moliere}
and Pacios \cite{Pacios}.
The Pacios approximation provides basically the same results
as frequently used Doyle-Turner scheme
\cite{DoyleTurner1968} but in contrast to the latter
it leads to the correct behaviour $\propto 1/r$ of
$U_{\rm at}$ at small distances (see, e.g., Ref. \cite{Dedkov1995}).
More details on the algorithms implemented in
\textsc{MBN Explorer} are given
\cite{MBN_Explorer_2012,KorolSushkoSolovyov:EPJD_v75_p107_2021}.

In the current paper we consider propagation of
855 MeV electrons through crystalline and amorphous
Silicon targets the parameters of which chosen in accordance
with the experiment carried out at the MAMI
facility \cite{MazzolariEtAl:EPJC_v80_63_2020}.
In the case of the oriented crystalline target the simulations
have been performed for the electron beam incident on the
crystal along the directions probed experimentally as well as
along the directions that were not explored in the experimental
setup.
This allows us to compare the simulated dependencies with the
experimentally measured data on the angular distribution of
deflected electrons as well as to investigate the evolution
of the distribution with variation of the relative orientation
of the crystal and the beam.
The atomistic approach implemented in \textsc{MBN Explorer}
enables simulation the trajectories of charged particles
entering a crystal along any chosen direction.
Taking advantage of this capability, we have specifically studied
the beam incident on a crystal at some angles with respect to
the $\left\langle 001\right\rangle$
crystallographic axis.

In Ref. \cite{MazzolariEtAl:EPJC_v80_63_2020} the directions
of the incident 855 MeV beam that lead to CSS have been specified
in terms of the planar angles  $\Theta_{\rm v}$, $\Theta_{\rm h}$
that can be used instead of the polar angles
$\Theta=\left(\Theta_{\rm v}^2+\Theta_{\rm h}^2\right)$,
$\phi=\tan^{-1}\left(\Theta_{\rm v}/\Theta_{\rm h}\right)$ measured with respect to the $\left\langle 001\right\rangle$ direction.
These angles are defined as follows:
(i) $\Theta_{\rm v}$ is the angle between ${\vec p}_0$  and the $(110)$
plane,
and
(ii) $\Theta_{\rm h}$ is the angle between ${\vec p}_0$ and the
$(1\bar{1}0)$ plane.
It was estimated that the angles' values are to be chosen
from the intervals (in mrad):
\begin{equation}
18  < \Theta_{\rm h} < 43\,,
\qquad
2   < \Theta_{\rm v} < 4.3 \,.
\label{eq:03}
\end{equation}

The experiment was carried out using for a 34.2 $\mu$m thick
silicon crystal.
The CSS effect in the angular distribution of the deflected
electrons was observed for $\Theta_{\rm h} = 34.9$ mrad and
$\Theta_{\rm v} = 3$ mrad.
The data obtained was compared to the angular distribution measured in  32.76 $\mu$m thick amorphous silicon.

In our simulations other values of
$\Theta_{\rm h}$ and $\Theta_{\rm v}$, that are close to this direction, have been considered.
To reproduce the experimental results \cite{MazzolariEtAl:EPJC_v80_63_2020} we have studied
the effect of multiple scattering suppression that takes
place at certain orientations of the crystalline target.
The goal was to determine the necessary conditions for CSS
to occur, which typically happens at the incident angle much
greater than Lindhard's critical angle, $\Theta_{\rm L}$.
Written in terms of the depth $U_0$ of the continuous
interplanar potential $\Theta_{\rm L}$ reads
$\Theta_{\rm L}=(2U_0/\E)^{1/2}$.
For a 855 MeV projectile in a (110) channel in a silicon
crystal ($U_0\approx 20$ eV) this estimate provides
$\Theta_{\rm L}\approx 0.22$  mrad.

In the Section below we present and discuss the results of
simulations.
Basing on the information provided on the experimental
setup, the simulations have been carried out for several
values of the angles $\Theta_{\rm h}$ and $\Theta_{\rm v}$
within the intervals specified by Eq. (\ref{eq:03}).
Four sets of the simulations have been performed for the
crystalline target.
Two sets refer to the fixed value of $\Theta_{\rm v}$
(2 and 3 mrad, correspondingly) and $\Theta_{\rm h}$
varied within the specified interval.
In another pair of sets the value of $\Theta_{\rm v}$ was
scanned over the interval while $\Theta_{\rm h}$ was
fixed  at 33 and 34.9 mrad.

For each pair $(\Theta_{\rm v},\Theta_{\rm h})$ considered
a large number, $N\approx 4\times 10^4$,
of trajectories were simulated.
In the course of the simulations the transverse
velocity of the particles at the solid target entrance were
generated taking into account
the Gaussian distributions of the beam particle due to
the beam divergencies.
The following values of the standard deviations of the
divergences of the 855 MeV electron beam at MAMI were used:
$\sigma_{\rm h}^{\prime} = 70$
and $\sigma_{\rm v}^{\prime} = 30$ $\mu$rad.
General methodology implemented in \textsc{MBN Explorer}
to generate particles' trajectories in an atomic environment
accounts for randomness in sampling the incoming projectiles
(the key factors here are beam size, divergence and direction
with respect to the crystal orientation) as well as in
displacement of the lattice atoms from the nodal positions
due to the thermal vibrations (see Refs.
\cite{MBN_ChannelingPaper_2013,KorolSushkoSolovyov:EPJD_v75_p107_2021}
for details).
As a result, each trajectory corresponds to a unique atomic
environment and, therefore, all simulated trajectories are
statistically independent and can be analyzed further to
quantify the process of interest, the multiple scattering
process in particular.

For the sake of comparison,
the same number of trajectories have been simulated
for 855 MeV electrons passing through amorphous silicon.

To simulate the amorphous environment the following
procedure was used.
In \textsc{MBN Explorer}, when constructing a crystalline
structure the position vectors of atomic nuclei are
generated with account for random displacement from the
nodes due to thermal vibrations corresponding to given
temperature $T$.
By introducing unrealistically large value of
the root-mean-square thermal vibration amplitude $u_T$
(for example, comparable to the lattice constant $a$)
it is possible to consider large random displacements, i.e.
the structure will resemble an amorphous medium.
Additionally, one can change the value of the
unit cell volume to make the volume density of atoms
equal to the volume density in the amorphous medium.
To simulate amorphous silicon medium $u_T$ was set to
2 \AA.
Taking into account that a unit cell of a
silicon crystal (with $a=5.43$ \AA) contains 8 atoms,
the quoted value of $u_T$ is sufficient enough to ensure
randomness of atomic positions in the sample.

%%%%%%%%%%%%%%%%%%%%%%%%%%%%%%%%%%%%%%
\section{Results and discussions\label{Results}}

The simulated trajectories have been statistically
analyzed to provide quantitative data on the particles'
distribution in the following two planar
angles:\footnote{In Ref. \cite{MazzolariEtAl:EPJC_v80_63_2020} the notations $\vartheta_x,\vartheta_y$ are 
used for the angles $\vartheta_{\rm v}, \vartheta_{\rm h}$}
\\
(i) $\vartheta_{\rm v}$ --  the angle between projectile's
momentum $\vec p$  at the target's exit and
the (110) plane;
\\
(ii) $\vartheta_{\rm h}$ --  the angle between
$\vec p$ and the $(1\bar{1}0)$ plane.
\\
These planar angles are measured in two perpendicular planes,
therefore, the total (round) scattering angle $\theta$ with respect to
the $\langle 001\rangle$ axis is calculated as
$\theta=\left(\vartheta_{\rm v}^2 + \vartheta_{\rm h}^2\right)^{1/2}$.

The distribution of deflected particles in either of the planar angles
can be derived from the distribution in $\theta$.
The latter, following Ref. \cite{MazzolariEtAl:EPJC_v80_63_2020}, we
consider within the framework of the Moli\`{e}re scattering model
\cite{Moliere} (see also a paper by Bethe \cite{Bethe-PR_v89_p1256_1953}).
The details of the derivation are presented in Appendix
\ref{MoliereDistribution}.
For a planar angle (in the formula below $\vartheta$ stands for either
of $\vartheta_{\rm v}$, $\vartheta_{\rm h}$) the probability distribution
function (PDF) that represents the angular distribution function of the
particles experienced multiple scattering, can be written as follows:
\begin{eqnarray}
\text{PDF}(\vartheta)
&=&
{1\over N}{\d N(\vartheta) \over \d \vartheta}
=
\calA \sum_{n=0}^{\infty}
\left({\chi_{\rm c}^2\over 2\theta_{\rm s}^2}\right)^n
{1\over n!} \times
\nonumber \\
&&
\hspace{-1cm}\times\int_0^{\infty}  \d u\,
\ee^{-u^2/4}
\cos\left(u {\vartheta\over \sqrt{2 \theta_{\rm s}^2}}\right)
\left({u^2\over 4} \ln{u^2\over4}\right)^n
\label{eq:04}
\end{eqnarray}
where $\d N(\vartheta)$ denotes a number of particles
that are scattered by a planar angle lying within
the interval $[\vartheta, \vartheta+\d\vartheta]$,
$N$ is the total number of particles,
$\calA = 1/\pi\left(2 \theta_s^2\right)^{1/2}$ is the normalization
factor, and parameter
$\chi_{\rm c}$, dependent on particle's energy and target's
thickness, volume density and atomic charge number, is defined
in Appendix \ref{MoliereDistribution}, Eq. (\ref{MD:eq.02}).
The parameter $\theta_{\rm s}$ is to be determined by fitting the PDF function to the
simulated data.
To fit the data generated in the simulations with the Moli\`{e}re PDF
the  Levenberg-Marquardt algorithm \cite{Levenberg-M,NumRec} was
implemented.
The term with $n=0$ corresponds to the normal (Gaussian) distributions
in which $\theta_{\rm s}$ plays the role of standard deviation.
The terms $n\geq1$, which provide the correction to the normal
distribution, can be effectively evaluated numerically (see Appendix
\ref{MoliereDistribution} for more details).
In the fitting procedure the series in  (\ref{eq:04}) were cut off at
$n=12$.

\begin{figure}[]
% \centering
\begin{center}
\includegraphics[width=8.5cm]{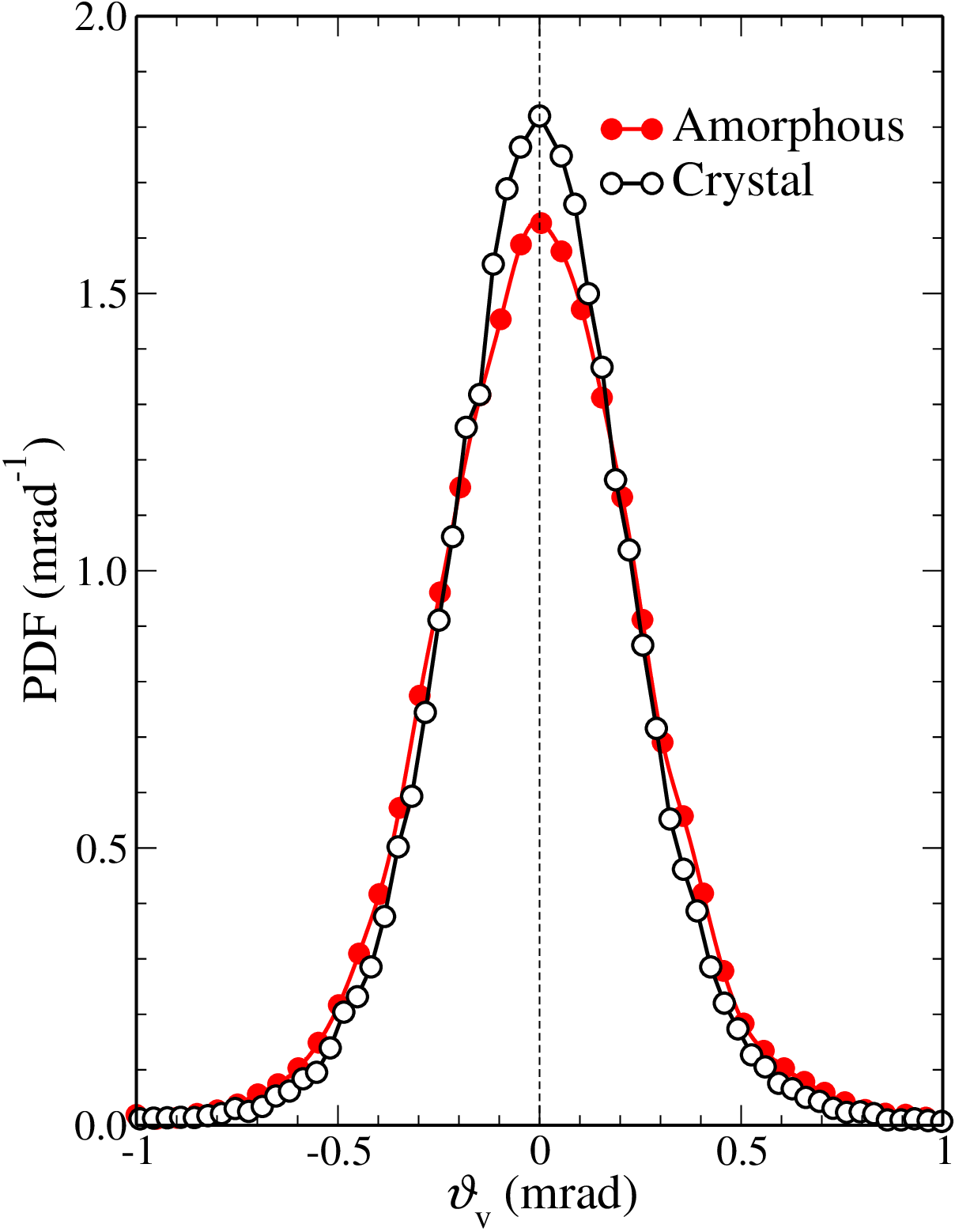}
\end{center}
\caption{
The probability density function (PDF) versus vertical scattering angle
$\vartheta_v$ plotted for amorphous (red line, filled circles) and
crystalline silicon (black line, open circles).
The data refer to the incident angles $\Theta_{\rm h} = 34.9$ mrad
and $\Theta_{\rm v}= 3$ mrad.
}
\label{Figure01.fig}
\end{figure}

The CSS effect, which was reported in the experiment, has also been
observed in our simulations within a specific region characterized by
$\Theta _{\rm h} = 34.9$ mrad and $\Theta_{\rm v}= 3$ mrad.
Figure \ref{Figure01.fig} compares the PDFs as functions of the vertical
scattering angle $\vartheta_{\rm v}$ plotted as histograms using the
data obtained from the simulations for the crystalline (black curve
with open circles) and amorphous (red curve with filled circles)
silicon targets.
It can be observed that the PDF values are higher for the crystal silicon
compared to the amorphous one in the domain off small scattering angles
$|\vartheta_{\rm v}| \lesssim 0.3$  mrad.
Notably, the peak value of the PDF obtained in the simulations is
$1.82$ mrad$^{-1}$ for the crystal which is slightly higher $1.67$
mrad$^{-1}$ for the amorphous target.
This suggests that the crystalline silicon exhibits a lower scattering
intensity within the specified range of scattering angles compared to the
amorphous state.
However, it is important to note that our simulation results show a
slight broadening of the calculated PDF compared to the experimentally
observed data.

%%%%%%%%%%%%%%%%%%

Following Ref. \cite{MazzolariEtAl:EPJC_v80_63_2020}, one can consider
the difference of PDFs for the crystal and amorphous Si,
PDF$_{\rm cr}$-PDF$_{\rm am}$, to quantify further the changes in the
angular distributions of electrons passing through the two types of medium.
This difference allows one to numerically evaluate the dissimilarity
in the PDFs for the two types of medium providing a quantitative measure
of the disparity in scattering behaviour as well as to  gain valuable
insights into the structural effects that influence
the electron multiple scattering.

\begin{figure}[]
\centering
\includegraphics[width=8.8cm]{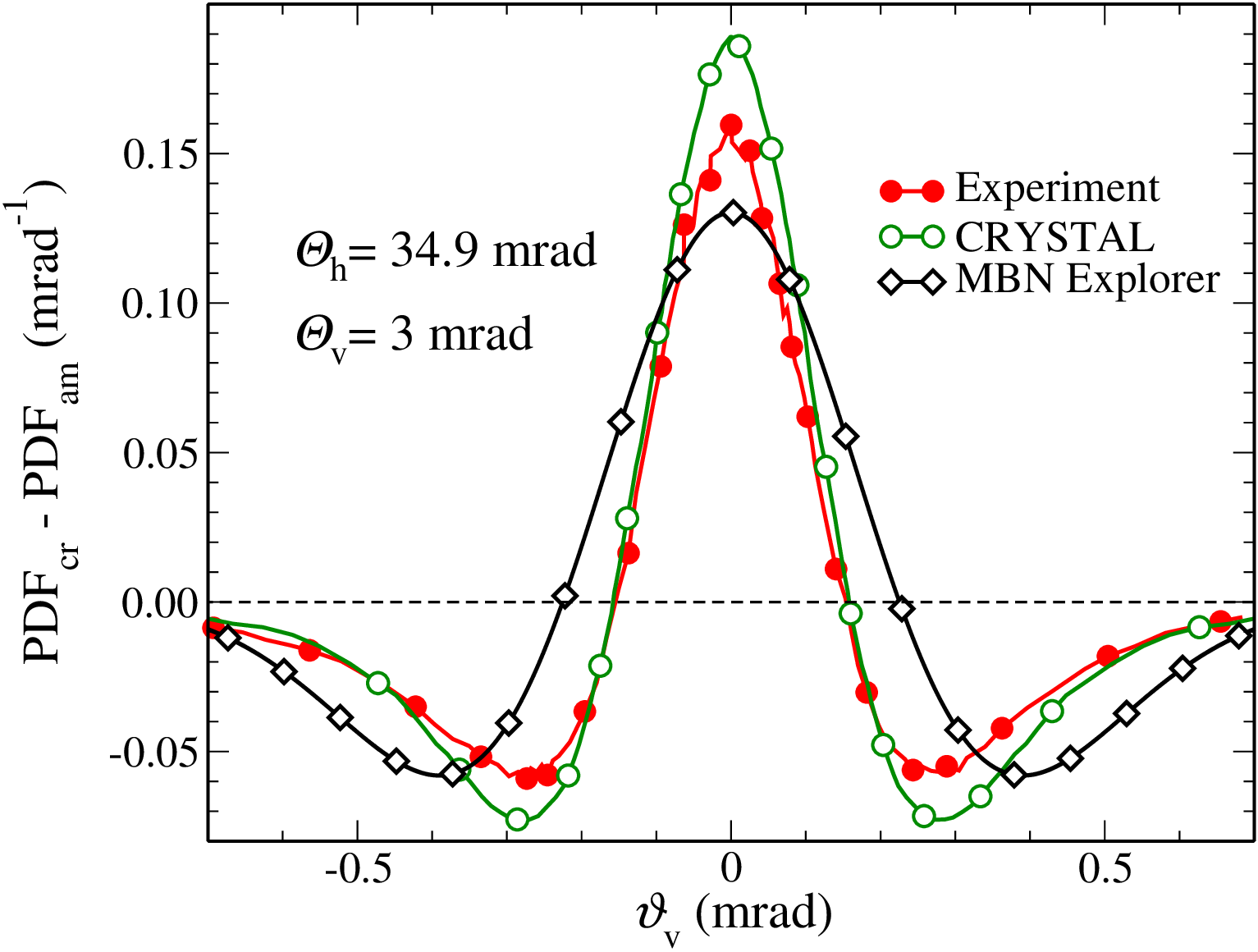}
\caption{
The difference of PDFs for the beam deflected by the silicon crystal
and the amorphous target.
Solid red line with filled circles and solid green line with open
circles stand for the
Experimental data (red line with filled circles) and
the results of the  CRYSTAL code (green line with open circles) are
from Ref. \cite{MazzolariEtAl:EPJC_v80_63_2020}.
Black line with open diamonds show the data simulated with
\textsc{MBN Explorer}.
In all cases the incident angles are $\Theta_{\rm v}= 3$ mrad and
$\Theta_{\rm h}= 34.9$ mrad.
}
\label{Figure02.fig}
\end{figure}

Figure \ref{Figure02.fig} shows the variations in the PDF for the beam
deflected by the crystal and amorphous targets.
The variations obtained by means of \textsc{MBN Explorer} (black curve
with diamonds) are compared to the experimental data (red line with
filled circles) and to the results of the simulations with the CRYSTAL
code \cite{SytovTikhomirov_NIMB_v355_p383_2015}
(green line with open circles) both taken from
Ref.\cite{MazzolariEtAl:EPJC_v80_63_2020}.
All data presented correspond to the incident angles
$\Theta_{\rm v} = 3$ mrad and
$\Theta_{\rm h} = 34.9$ mrad in the case of the crystalline target.
The results of the simulations performed with \textsc{MBN Explorer}
shown in the figure correspond to the Moli\`{e}re potential.
Similar simulations have been performed with the Pacios potential
(not shown in the figure) and it was that the differences between
the two sets of data is less than 0.4\%.
We consider this as a demonstration of the robustness of our results.
All three curves presented in the figure show similar behaviour.
However, there are some differences between the currents simulation and the experimental results regarding, in particular, the width of the central peak of the function.

 %%%%%%%%%%%%%%%%%%%%%%%%%%%%%%%%%

In our simulations we aimed to investigate the CSS effect over a wide range
of the incident angles.
Therefore, we have explored values beyond the optimal ranges indicated
in Eq. (\ref{eq:03}) to thoroughly understand the scattering behaviour and to assess the robustness of the observed CSS effect.

Following the approach adopted in the experiment
\cite{MazzolariEtAl:EPJC_v80_63_2020},
we have conducted scans over the incident angles aiming at exploring
the variation in the scattering behaviour of electrons in crystalline
silicon in comparison with the amorphous medium.
To this end, two sets of simulations have been carried out.
In the first set, the the vertical incident angle was fixed at
$\Theta_{\rm v}=3$ mrad whereas the horizontal angle
was varied within the interval $8.7\leq \Theta_{\rm h} \leq 52$ mrad.
Similarly, in the second set the angle $\Theta_{\rm h}=34.9$ mrad
was fixed and the angular distributions was explored in the range
$1 \leq\Theta_{\rm v}\leq 10$ mrad.
These scanning parameters allowed us to investigate the angular
dependence of the scattering phenomenon and to identify notable
trends or/and extrema of the PDFs.
For all combinations $(\Theta_{\rm h},\Theta_{\rm v})$ considered
the simulations were performed using both the Moli\`{e}re and
Pacios potentials.
The PDF obtained were practically independent on the potential used.
The minimum value of $\theta_{\rm s}$ due to the coherent scattering
has been found at $\Theta_{\rm h}=34.9$ mrad and $\Theta_{\rm v}=2$ mrad.
This observation provided a crucial starting point for further analysis
and understanding of the scattering behavior in silicon.
To quantify the scattering distribution and to derive the PDF for
scattered electrons additional scans were performed fixing
$\Theta_{\rm v}=2$ mrad and varying $\Theta_{\rm h}$ as well as
scanning over $\Theta_{\rm v}$ while fixing $\Theta_{\rm h}=33$ mrad.
These scans allowed us to capture the variations in the PDF as we explored different
scattering angles.

%%%%%%%%%%%%%%%%%%%%%%
 \begin{figure*}[ht!]
\centering
\includegraphics[width=8.5cm]{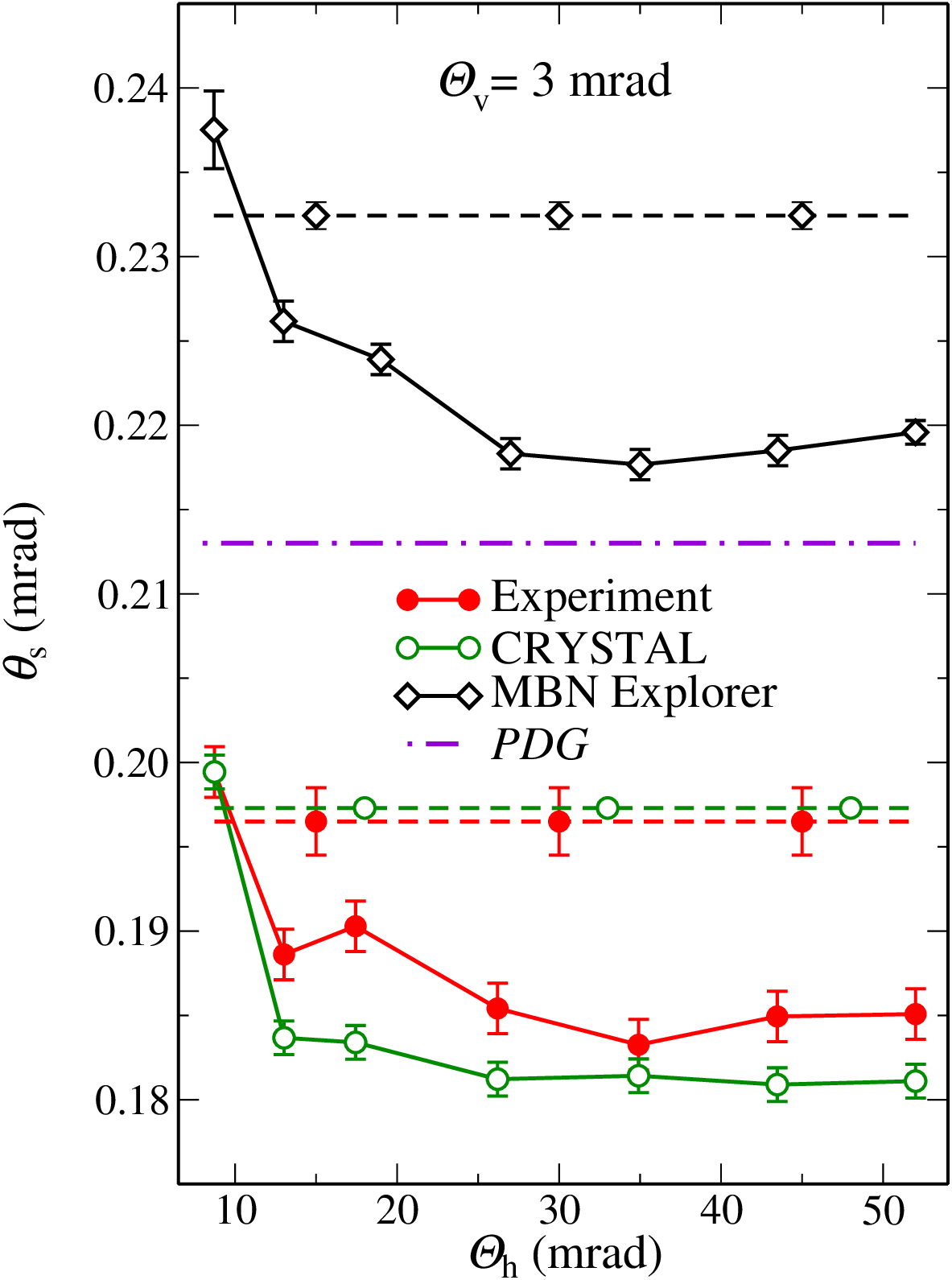}\hspace*{0.3cm}
\includegraphics[width=8.5cm]{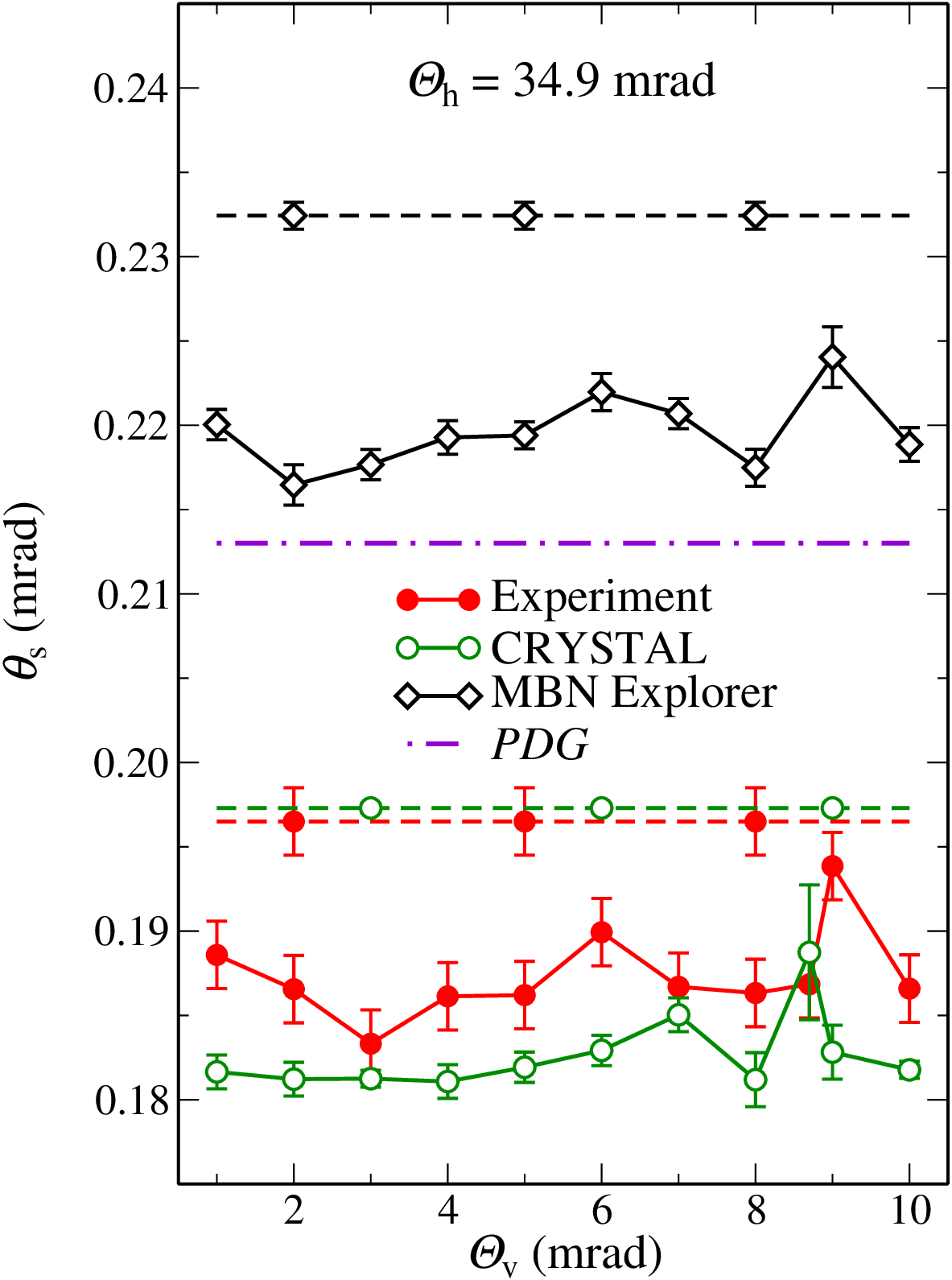}
\caption{
\textit{Left:}
Multiple scattering angle $\theta_{\rm s}$ as a function of
the horizontal
(left) and vertical (right) incident angles.
Solid lines stand for the crystalline target, dashed lines
indicate the results for the amorphous medium.
Experimental data (red lines, filled circles) and results
of simulations with the CRYSTAL code (green lines, open
circles) are from Ref. \cite{MazzolariEtAl:EPJC_v80_63_2020}.
The results of current simulations (black lines with diamonds)
correspond to the Moli\`ere potential.
For the crystal Si, the incident angles are fixed at:
$\Theta_{\rm v}=3$ mrad (left) and $\Theta_{\rm h}=34.9$ mrad
(right).
In both graphs dashed line with dots denotes the rms scattering angle
calculated using the formula presented in the review by
Particle Data Group (PDG) \cite{WorkmanEtAl:ParticleDataGroup_2022}.
}
\label{Figure03.fig}
\end{figure*}

Figure \ref{Figure03.fig} shows the variation of
$\theta_{\rm s}$ as a function of one of the incident angles
($\Theta_{\rm h}$ in left graph and $\Theta_{\rm v}$ in right graph) while fixing the second incident angle (see the values
indicated at the top of each graph).
Solid and dashed lines stand for the crystalline and the
amorphous targets, respectively.
The results of our simulations are shown as open
diamond symbols connected with black lines.
Statistical uncertainties due to the
finite number of simulated trajectories are indicated with
error bars.
The data reported in Ref. \cite{MazzolariEtAl:EPJC_v80_63_2020}
are shown with filled circles connected with red lines (the
experimental data) and with open circles connected with green lines
(the results of simulation with the CRYSTAL code
\cite{SytovTikhomirov_NIMB_v355_p383_2015}).
For the sake of comparison we also present the value of
multiple scattering angle calculated for the amorphous silicon by means of  Eqs. (34.16),
(34.25) and (34.26) from the review by Particle Data Group
\cite{WorkmanEtAl:ParticleDataGroup_2022} (see also
Appendix \ref{ParticleDataGroup}).
This value, shown in both graphs with dashed-dotted line, corresponds
to the mass thickness $T\propto NL$ ($N$ is the volume density,
$L$ is the thickness) that one can calculate from the information
on the samples used in the experiment indicated
Ref. \cite{MazzolariEtAl:EPJC_v80_63_2020}.
It is indicated that mass thicknesses of the crystalline and amorphous samples were equal, therefore,  $T\propto N_{\rm cr}L_{\rm cr}$ with
$L_{\rm cr}=32.76$ $\mu$m as quoted and
$N_{\rm cr}=8/(5.43\,\mbox{\AA})^3=5.00\times 10^{22}$ cm$^{-3}$.

One can state that there is a close similarity between our results and
the experimentally measured data when comparing the profiles of the
dependencies $\theta_{\rm s}\left(\Theta_{\rm h}\right)$ and
$\theta_{\rm s}\left(\Theta_{\rm v}\right)$
for the crystal target.
Indeed, the simulations have successfully captured all the
essential trends in the dependencies and reproducing the key
features observed in the experiment.

To be noted, nevertheless, is a systematic difference in absolute
values of the simulated and measured data on  $\theta_{\rm s}$.
For both crystalline and amorphous targets the simulations produce
higher values with the excess of ca 0.03-0.035 mrad on
average.\footnote{Also to be noted is that for the amorphous target
our result and the experimental data both differ from the commonly used
estimate \cite{WorkmanEtAl:ParticleDataGroup_2022} of the multiple scattering angle, see the dashed-dotted lines in both graphs in Fig.
\ref{Figure03.fig}.}
The reason for this discrepancy is unclear.
One can speculate that it might be related to some specific algorithm in
the statistical processing of the results.
For example, it was noted in Ref. \cite{MazzolariEtAl:EPJC_v80_63_2020}
that to compare directly the results obtained by means of the CRYSTAL
code to the experimental data the former are to be convoluted with
the angular distribution of the beam.
However, this is definitely not the issue in connection with our
methodology since the beam divergence has been explicitly accounted for
to randomize "entrance conditions" for the incident particles
(in both vertical and horizontal directions).
The feature that might be relevant but deserves further analysis
concerns the algorithm implemented in \textsc{MBN Explorer} to simulate
crystalline and amorphous environment.
As mentioned in the last paragraph in Section \ref{Theory}
(see \cite{KorolSushkoSolovyov:EPJD_v75_p107_2021,MBN_ChannelingPaper_2013}
for more details) the atomic displacements from the nodal positions
are randomly generated.
For doing this the normal distribution with the variance equal to $u_T$
is used.
This algorithm does not exclude non-physical events when two (or more) atoms will be located in close vicinity to each other.
This will result in the (local) increase in the scattering angle which
inevitably will affect the multiple scattering angle as well.

%%%%%%%%%%%%%%%%%%%%%
\section{Conclusion \label{Conclusion}}

By means of relativistic all-atom molecular dynamics implemented
in the \textsc{MBN Explorer} software package the passage
of ultra-relativistic electrons through crystalline and
amorphous silicon media has been simulated.
The parameters of the 855 MeV electron beam and the thickness
of the silicon targets utilized in the simulations correspond
to those used in the experiment  \cite{MazzolariEtAl:EPJC_v80_63_2020}
which revealed broad angular anisotropy of multiple scattering in the crystal.
For each incident geometry considered a sufficiently large number
($\approx 4\times 10^4$)
of statistically independent trajectories has been generated.
The statistical analysis carried out resulted in calculation of
the probability distribution functions (PDF) that allowed us
to quantify the process of multiple
scattering in terms of the planar scattering angles.
To double check the results, the simulations and the analysis have been
performed employing both the Moli\`{e}re and Pacios potentials for the
electron--Si-atom interaction.
The dependencies of the scattering angle $\theta_{\rm s}$ on both the
horizontal $\Theta_{\rm h}$ and vertical $\Theta_{\rm v}$ incident
angles turned out to be virtually insensitive to the choice of the
potential.

We have observed the differences in angular distributions of electrons
deflected by the crystalline and the amorphous targets that were
predicted, explained and measured in Ref. \cite{MazzolariEtAl:EPJC_v80_63_2020}.
Our results exhibit a high degree of agreement with the experimental data
in terms of the profiles of the dependencies
$\theta_{\rm s}\left(\Theta_{\rm h}\right)$ and
$\theta_{\rm s}\left(\Theta_{\rm v}\right)$.
We have successfully reproduced values of $\theta_{\rm s}$ for which
the phenomenon of coherent scattering suppression was detected and
quantified experimentally.
In addition, we have examined this phenomenon over a broad range
of the incident beam directions.

In terms of absolute values of the scattering angle
$\theta_{\rm s}$ there
is a systematic difference between the simulated and experimentally
measured data.
The simulations produce the values of $\theta_{\rm s}$ that are
approximately 30-35 $\mu$rad higher.
As of now, we are uncertain on the origin of this systematic
discrepancy.
It might be related difference in the methods of statistical analysis
adopted in the current work and in Ref.
\cite{MazzolariEtAl:EPJC_v80_63_2020}.
We are planning to carry out a study whether this
feature might originate from the algorithms implemented in
\textsc{MBN Explorer} to simulate crystalline and amorphous media.
However, we also believe that further experimental studies of the phenomenon are needed.

%%%%%%%%%%%%%%%%%%%%%%%%%
\section{Acknowledgements}

We acknowledge support by the European Commission
through the N-LIGHT Project within
the H2020-MSCA-RISE-2019 call (GA 872196)
and the EIC Pathfinder Pro\-ject TECHNO-CLS
(Project No. 101046458).
GRL, MMM and JRS would like to thank NA223LH-INSTEC-003 pro\-ject
from InSTEC-UH.
We also acknowledge the Frankfurt Center for Scientific
Computing (CSC) for providing computer facilities.

We are grateful to Andrea Mazzolari, Laura Bandiera and
Alexey Sytov for the communications that allowed us to
clarify some experimental and computational issues raised
in connection with Ref. \cite{MazzolariEtAl:EPJC_v80_63_2020}

%%%%%%%%%%%%%%%%%%%%%%%%%%%%%%%
\section{Authors contributions}
All the authors were involved in the preparation of the
manuscript and contributed equally to this work.
All the authors have read and approved the final manuscript.

%%%%%%%%%%%%%%%%%%%%%%
\appendix\markboth{Appendix}{}
\renewcommand{\thesection}{Appendix \Alph{section}}
\renewcommand{\thesection}{\Alph{section}}
\numberwithin{equation}{section}

%%%%%%%%%%%%%
\section{Basic formulae related to
the Moli\`ere distribution \label{MoliereDistribution}}

In what follows we outline the derivation of the 1D
distribution
function $\d P(\vartheta)/\d \vartheta$ with
respect to a planar scattering angle $\vartheta$,
Eq. (\ref{eq:04}), starting with the Moli\`ere 2D distribution
function \cite{Moliere} written
in terms of the solid scattering angle, $\d^2 P(\Om)/\d \Om$,
where $\d\Om = \theta \d\theta \d \phi$ with $\theta\ll 1$ and $\phi$
being the polar angles.
Also presented are auxiliary formulae related to
the numerical evaluation of the fitting parameter
$\theta_{\rm s}$ by means of the Levenberg-Marquardt algorithm
\cite{Levenberg-M,NumRec}.

When referring to the Moli\`ere distribution formalism
the notations introduced in Ref.
\cite{Bethe-PR_v89_p1256_1953} are used.

Introducing planar scattering angles
$\vartheta_x=\theta\cos\phi$ and
$\vartheta_y=\theta\sin\phi$, so
that $\theta = \left(\vartheta_x^2 + \vartheta_y^2\right)^{1/2}$
and $\d\Om = \d\vartheta_x \d\vartheta_y$, one writes
the 2D distribution as follows:
\begin{eqnarray}
{\d^2 P(\vartheta_x , \vartheta_y) \over \d\vartheta_x \d\vartheta_y}
=
A
\sum_{n=0}^{\infty}
\left({\chi_c^2\over 2\theta_s^2 }\right)^n
f^{(n)}(\vartheta_x , \vartheta_y)\,.
\label{MD:eq.01}  %{theta_xy:eq.05}\\
\end{eqnarray}
Here
\begin{eqnarray}
\chi_c^2
\approx
4\pi N L { e^4 Z(Z+1) z^2 \over \E^2}
\label{MD:eq.02}
\end{eqnarray}
where $z, p, v, \E\approx pv$ are, respectively,
charge of the projectile (in units of
the elementary charge $e$), its momentum, speed and
energy calculated in the ultra-relativistic limit;
$N$ is the volume density (in 1/cm$^3$) of atoms in the medium,
$Z$ is their atomic number,
$L$ stands for the target's thickness.
Physical meaning of $\chi_c$ is that the total
probability of a single scattering
through an angle greater than
$\chi_c$  is exactly one.

The functions $f^{(n)}(\vartheta_x , \vartheta_y)$ are given by
\begin{eqnarray}
f^{(n)}(\vartheta_x , \vartheta_y)
&=&
{1\over n!}
\int_0^{\infty}
J_0\left(u\sqrt{{\vartheta_x^2 + \vartheta_y^2 \over 2 \theta_s^2}}
\right)
\ee^{-u^2/4}
\nonumber\\
&\times&
\left({u^2\over 4} \ln{u^2\over4}\right)^n u \d u\,.
\label{MD:eq.03}
\end{eqnarray}

The normalization factor $A$ can be calculated explicitly if both scattering
angles $\vartheta_{x,y}$ vary within the infinite interval $[-\infty,+\infty]$.
The result reads $A=1/4\pi \theta_s^2$.

The planar angle $\theta_{\rm s}$, which enters Eqs. (\ref{MD:eq.01}) and
(\ref{MD:eq.03}),  is the only free parameter.
It is to determined by fitting the simulated or experimentally acquired data with
the Moli\`ere distribution function.

To derive the 1D distribution one
integrates Eq. (\ref{MD:eq.01})
with respect to either variable $\vartheta_{x,y}$ over the infinite interval.
Then, using the relation (see, e.g., Eq. 6.554.3 in Ref. \cite{Gradshteyn})
\begin{eqnarray}
\int_{-\infty}^{\infty}
J_0\left(a\sqrt{x^2 + b^2} \right)
\d x
=
2
{\cos\left(ab \right) \over a},
\label{MD:eq.04}
\end{eqnarray}
one writes the normalized Moli\`ere partial distribution function (PDF)
with respect to a single planar
angle $\vartheta$ as follows
\begin{eqnarray}
\mathrm{PDF}(\vartheta; \theta_s)
= {\d P(\vartheta) \over \d\vartheta}
=
\calA
\sum_{n=0}^{\infty}
\calB^n
\calC^{(n)}(\xi)
\label{MD:eq.05}
\\
\calC^{(n)}(\xi)
=
{1\over n!}
\int_0^{\infty}  \d u\,
\ee^{-u^2/4}
\cos(u \xi)
\left({u^2\over 4} \ln{u^2\over4}\right)^n
\label{MD:eq.06}
\end{eqnarray}
where short-hand notations have been introduced:
\begin{eqnarray}
\calA = {1\over \pi\sqrt{2 \theta_s^2}},
\qquad
\calB = {\chi_c^2\over 2\theta_s^2 },
\qquad
\xi={\vartheta \over \sqrt{2 \theta_s^2}}\,.
\label{MD:eq.07}
\end{eqnarray}
Note that $\mathrm{PDF}$ depends parametrically on $\theta_s$.
This dependence appears explicitly in the normalization factor $\calA$,
in the factors $\calB^n$ and in the argument $\xi$.

For $n=0$ the integral is carried out
explicitly,\footnote{Explicit formula for
$\calC^{(1)}$ is also known,
see Refs. \cite{Moliere,Bethe-PR_v89_p1256_1953}.}
$\calC^{(0)}(\xi)=\sqrt{\pi}\exp\left(-\xi^2\right)$.
Therefore, the first term in the series (\ref{MD:eq.05})
corresponds to the Gaussian distribution with variance
$\theta_s$.
Efficient numerical evaluation of the integrals $\calC^{(n)}$
with $n\geq 1$ can be achieved by means
of the Gauss-Laguerre quadratures (see, e.g., Ref.\cite{NumRec}).

Fitting the data with the distribution (\ref{MD:eq.05}) implies
establishing the value of the only free parameter, $\theta_s$, of the
distribution.
To determine its value one can use the Levenberg-Marquardt algorithm \cite{Levenberg-M,NumRec}.
To implement this algorithm an explicit expression for the PDF derivative
with respect to $\theta_s$ must be known.
For the sake of completeness, this formula is written below:
\begin{eqnarray}
\!\!\!\!
{\d \mathrm{PDF} \over \d\theta_s}
=\!
-
{1 \over \theta_s}
\left[
\mathrm{PDF}
+
\calA
\sum_{n=0}^{\infty}
\calB^{n}\!
\left(
2n\calC^{(n)}
+
\xi\calD^{(n)}
\right)
\right]
\label{my:eq.07a}
\end{eqnarray}
where
\begin{eqnarray}
\calD^{(n)}
\!=\!
{\d \calC^{(n)} \over \d\xi}
\!=\!
-{1\over n!}
\int\limits_0^{\infty}\!\!\! u\, \d u
\ee^{-u^2/4}
\sin(u \xi)\!
\left({u^2\over 4} \!\ln{u^2\over4}\right)^n
\label{my:eq.03}
\end{eqnarray}

%%%%%%%%%%%%%
\section{Particle Data Group \label{ParticleDataGroup}}

A planar rms multiple scattering angle
$\theta_0$
for an ultra-relativistic electron / positron ($pv\approx \E$)
moving in an amorphous medium of density $\rho_{\rm am}$ and
thickness $L_{\rm am}$ can be calculated as follows
(see Ref. \cite{WorkmanEtAl:ParticleDataGroup_2022}, Eq. (34.16)):
\begin{eqnarray}
\theta_0
=
{13.6 \, \mbox{[MeV]} \over \E}
\sqrt{\rho_{\rm am} L_{\rm am} \over X_0}\!\!
\left(1 + 0.038\ln{\rho_{\rm am} L_{\rm am} \over X_0}\right).
\label{PDG:eq.01}
\end{eqnarray}
Here $X_0$ is the medium's radiation length (measured in g/cm$^2$)
that determines the mean distance over which a high-energy electron loses
all but $1/e$ of its energy by bremsstrahlung, see
Eq. (34.25) in \cite{WorkmanEtAl:ParticleDataGroup_2022}.

%%%%%%%%%%%%%%%%%%%%%%%%%%

\end{document}